\documentclass[fleqn,usenatbib]{mnras}

\usepackage{newtxtext,newtxmath}

\usepackage[T1]{fontenc}

\DeclareRobustCommand{\VAN}[3]{#2}
\let\VANthebibliography\thebibliography
\def\thebibliography{\DeclareRobustCommand{\VAN}[3]{##3}\VANthebibliography}

\usepackage{graphicx}	
\usepackage{amsmath}	
\usepackage{orcidlink}

\newif\ifpaper
\papertrue
\ifpaper
\fi

\title[Timing of solar flare footpoints]{Precise timing of solar flare footpoint sources from mid-infrared observations}

\author[P. J. A. Sim\~oes et al.]{
Paulo J. A. Sim\~oes$^{1,2}$\thanks{E-mail: paulo@craam.mackenzie.br}\orcidlink{0000-0002-4819-1884},
Lyndsay Fletcher$^{2,3}$ \orcidlink{0000-0001-9315-7899},
Hugh S. Hudson$^{2,4}$ \orcidlink{0000-0001-5685-1283},
Graham S. Kerr$^{5,6}$ \orcidlink{0000-0001-5316-914X},
\newauthor
Matt Penn$^{7}$, 
Karla F. Lopez$^{1}$ \orcidlink{0000-0002-3661-3517}
\\
$^{1}$Centro de R\'adio Astronomia e Astrof\'isica Mackenzie, Escola de Engenharia, Universidade Presbiteriana Mackenzie, 01302-907 S\~ao Paulo, Brazil.\\
$^{2}$SUPA School of Physics and Astronomy,University of Glasgow, Glasgow, G12 8QQ, UK \\
$^{3}$Rosseland Centre for Solar Physics, University of Oslo, PO Box 1029 Blindern, NO-0315 Oslo, Norway\\
$^{4}$Space Sciences Laboratory, University of California, Berkeley, CA, USA\\
$^{5}$NASA Goddard Space Flight Center, Heliophysics Science Division, Code 671, 8800 Greenbelt Rd., Greenbelt, MD 20771, USA \\
$^{6}$Department of Physics, Catholic University of America, 620 Michigan Avenue, Northeast, Washington, DC 20064, USA\\
$^{7}$Department of Physics, Southern Illinois University Carbondale, IL 62901, USA
}

\date{Accepted XXX. Received YYY; in original form ZZZ}

\pubyear{2024}

\begin{document}
\label{firstpage}
\pagerange{\pageref{firstpage}--\pageref{lastpage}}
\maketitle

\begin{abstract}
Solar flares are powerful particle accelerators, and in the accepted standard flare model most of the flare energy is transported from a coronal energy-release region by accelerated electrons which stop collisionally in the chromosphere, heating and ionising the plasma, producing a broadband enhancement to the solar radiative output. We present a time-delay analysis of the infrared emission from two chromospheric sources in the flare SOL2014-09-24T17:50 taken at the McMath-Pierce telescope. By cross-correlating the intensity signals, measured with 1s cadence, from the two spatially resolved infrared sources we find a delay of $0.75\pm0.07$~s at $8.2\mu$m, where the uncertainties are quantified by a Monte Carlo analysis. The sources correlate well in brightness but have a time lag larger than can be reasonably explained by the energy transport dominated by non-thermal electrons precipitating from a single acceleration site in the corona. If interpreted as a time-of-flight difference between electrons traveling to each footpoint, we estimate time delays between $0.14$~s and $0.42$~s, for a reconnection site at the interior quasi-separatrix layer or at the null-point of the spine-fan topology inferred for this event. We employed modelling of electron transport via time-dependent Fokker-Planck and radiative hydrodynamic simulations to evaluate other possible sources of time-delay in the generation of the IR emission, such as differing ionisation timescales under different chromospheric conditions. Our results demonstrate that they are also unable to account for this discrepancy. This flare appears to require energy transport by some means other than electron beams originating in the corona.

\end{abstract}




\section{Introduction}

Solar flares occur when magnetic energy stored in the corona rapidly converts into other forms and is transported to the lower solar atmosphere (the chromosphere), leading to the strong heating and intense radiation that characterises the flare.
In this paper we present a new analysis of flares in the infrared (IR) that is in clear contradiction to the long-standing and dominant model for flare energy transport from the corona to the chromosphere, the ``thick target'' model invoking electron beams \citep{EmslieBeam1978ApJ...224..241E,1972SoPh...24..414H}, a concept widely employed for flare simulations \citep[e.g.][]{Fisher1985ApJ...289..425F,2005ApJ...630..573A,2015ApJ...809..104A,Kowalski2017ApJ...836...12K,kong2022ApJ...941L..22K}. Our conclusion is possible due to the excellent signal-to-noise ratio and high time resolution of the new mid-IR observations, and {supported by} modeling of electron transport and IR emission in an evolving flare atmosphere.

In the typical solar flare model, electrons accelerated from the background thermal distribution in the solar corona directly tap the stored magnetic energy there and, moving along the field at a substantial fraction of the speed of light, \textit{de facto} transport this energy to the chromosphere where it is ultimately deposited via 
Coulomb collisions. This results in near-instantaneous production of non-thermal hard X-ray (HXR) bremsstrahlung radiation with energies of tens of keV, resulting in the appearance of chromospheric footpoints, accompanied by rapid plasma heating. The existence of accelerated coronal electrons is not in doubt; this is clear from the presence of coronal radio and HXR emission  \citep{1985SoPh..100..537L,Fletcher2011SSRv..159...19F,Holman2011SSRv..159..107H,Kontar2011SSRv..159..301K}. These do not, however, establish a dominant role for electrons in flare energy transport, which requires examination of the chromospheric signatures \citep{Battaglia2009A&A...498..891B,battaglia2014ApJ...789...47B,hudson2021MNRAS.501.1273H,lopez2022A&A...657A..51L,kerr2023FrASS...960862K,douglas2023MNRAS.525.4143D}. 

Footpoints that are magnetically linked to each other and to the same energy release site are known as conjugate footpoints. HXR conjugate footpoints often occur in pairs that appear to be at either end of a coronal flare loop, but more complex topologies are also inferred. 
Given the typical length of coronal flare loops, and separation of flare footpoints in the lower atmosphere, electrons traveling through the corona at 0.3~c (30~keV) would be expected to produce near-simultaneous correlated emission in the conjugate HXR sources frequently seen. 
Evidence for simultaneity was sought in the 1990s using images obtained at a cadence of 2~s from the \textit{Yohkoh} Hard X-ray Telescope (HXT) \citep{1994PhDT.......335S,1998opaf.conf..273S}. However, HXT's Fourier synthesis imaging has low signal-to-noise ratio and poor \textit{u,v}-plane sampling \citep{kosugi1992PASJ...44L..45K,Sato1999PASJ...51..127S}. This was compounded by the imaging reconstruction methods used, which give the smoothest image consistent with the data but potentially move counts between sources compromising the reported near-simultaneity. 

The discovery of an energy-dependent time delay between HXR lightcurves suggests the ``time-of-flight'' effects expected in the standard beam model \citep[][\textit{et seq.}]{1995ApJ...447..923A}, but the analysis does not unambiguously establish the model \citep{1998ApJ...509..911B}.
 
Analysis of conjugate footpoints using ultraviolet (UV) direct imaging has shown several instances of time lags between footpoint pairs significantly in excess of a second \citep{2009A&A...493..241F}. However, these 2~s cadence UV data were filtered to emphasise timescales of order 10~s, so spurious signals could have been introduced. Additionally, chromospheric UV emission can be excited by rather weak energy fluxes, and does not necessarily reflect energetically important parts of the flare \citep{2001ApJ...560L..87W}.

These HXR and UV results highlight the need to examine footpoint timing using high cadence, direct imaging of chromospheric radiation that is closely associated with the energetically dominant processes in a flare. Suitable HXR observations with sufficient resolution do not exist, and so here we use observations made at 1~s cadence in the mid-infrared, at the National Solar Observatory's McMath-Pierce telescope \citep{2016ApJ...819L..30P} at 5.2 $\mu$m and 8.2 $\mu$m ($\approx$ 57.7~THz and $\approx$ 36.6~THz, respectively). The sensors used from the McMath observations now have moved to Big Bear Solar Observatory to become its MIRI (Mid-IR Instrument) there.
Flare infrared sources have strong associations with HXR footpoints in space and in time \citep{2004ApJ...607L.131X,2013ApJ...768..134K}, and have source timing precision significantly better than any other observational signature to date \citep{1994PhDT.......335S,1995ApJ...447..923A}. They also have high contrast against the non-flaring chromosphere, so that detectability is excellent. 

This article is organized as follows: in Section~\ref{sec:data} we present an overview of the event,  a multi-wavelength data analysis and measurements of the time delay between the two mid-IR sources; in Section~\ref{sec:geom} we discuss the magnetic field geometry and spine-fan topology, and its implications for the electron transport timing;  
in Section~\ref{sec:transport} we employ time-dependent kinetic simulations for the transport of the non-thermal electrons and in Section~\ref{sec:radyn} we investigate the timescales for the formation of the mid-IR emission using radiative-hydrodynamic simulations, to evaluate possible sources of the observed time delay. Lastly, we present our conclusions and final remarks in Section~\ref{sec:conclusion}.

\section{Data Analysis}
\label{sec:data}
\subsection{Event Overview} 

\begin{figure*}[h]
\centering
  \includegraphics[width=\textwidth]{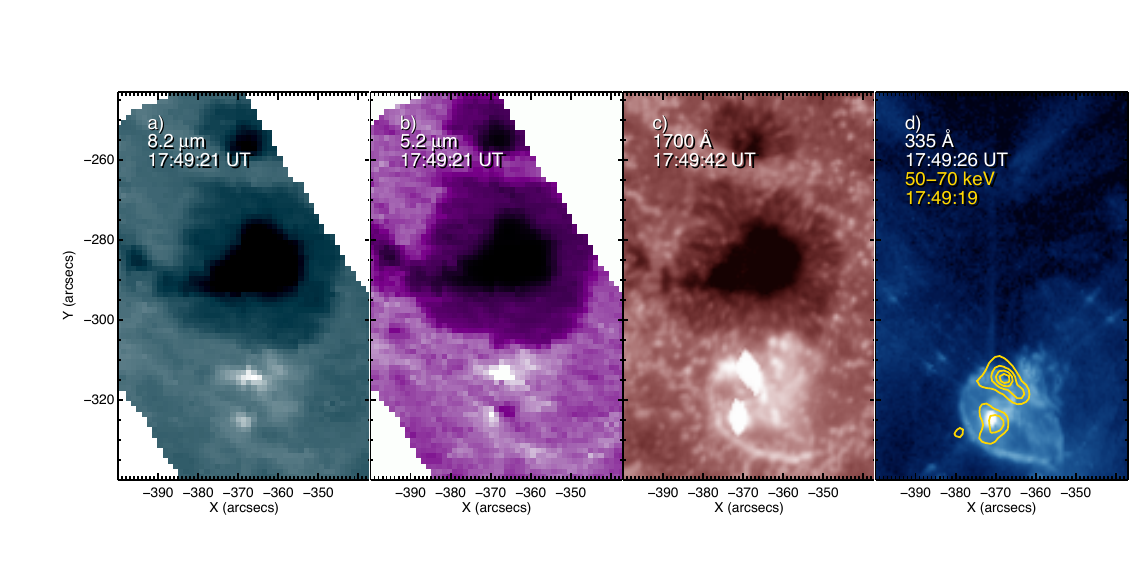}
        \caption{{Images of SOL2014-09-24 near the peak of the impulsive phase}, (a) IR 8.2~$\mu$m. (b) IR 5.2~$\mu$m. (b) SDO/AIA UV 1700 \AA\ (d) SDO/AIA EUV 335 \AA, overlaid with a RHESSI CLEAN image in the energy range 50--70 keV, with the contour levels at 25\%, 45\%, 65\%, and 85\% of the maximum of the emission. {The flare footpoints are evident as the compact bright patches seen in each wavelength.} 
        \label{fig:mcmp}
        }
\end{figure*}

The observations of SOL2014-09-24T17:50 \citep{2016ApJ...819L..30P} represent a breakthrough in signal-to-noise ratio and sampling for impulsive-phase signatures of flares in the lower solar atmosphere. Fig.~\ref{fig:mcmp} shows the variety of data available for this flare in the IR, (extreme) ultraviolet (EUV) from the \textit{Atmospheric Imaging Assembly} \citep[AIA,][]{AIA2012SoPh..275...17L}, on board the \textit{Solar Dynamics Observatory} \citep[SDO,][]{pesnell_2012SoPh..275....3P}, and HXR from the \textit{Reuven Ramaty High Energy Solar Spectroscopic Imager} \citep[RHESSI,][]{LinDennisHurford:2002}.
The flare's multi-wavelength emissions span the rapidly-evolving atmosphere from photosphere (optical) to upper chromosphere (EUV), and the presence of flare impulsive-phase signatures in the mid-IR supports the idea that broadband heating and excitation occur deep in the atmosphere \citep{1993ApJ...416L..91M,1994ApJ...422L..25H}. The formation of this continuum requires rapid increases of the ionization level at heights at least down to roughly 1000~km as a result of intense energy deposition \citep{2013ApJ...768..134K,2015SoPh..290.2809T,2016ApJ...819L..30P,simoes2017A&A...605A.125S}.

Fig.~\ref{fig:mcmp} shows the flare morphology at the time of the maximum emission in the IR, at 17:49:21 UT. Images at 8.2 and 5.2~$\mu$m (Fig.~\ref{fig:mcmp}a-b) show two compact bright patches (northern and southern footpoints; NF and SF), near a sunspot. These are the focus of this study and further discussed in Section~\ref{sec:IRdata}. Observations from SDO/AIA reveal similar compact footpoints (1700\;\AA\ and 335\;\AA\ images are shown in Fig.~\ref{fig:mcmp}c-d; note some small saturation artifacts in 1700 \AA). In Fig.~\ref{fig:mcmp}d the footpoints are also easily visible in hard X-rays reconstructed images from RHESSI, in the range of energy 50-70 keV (more details are presented in Section~\ref{sec:hxr}).

\begin{figure*}
\centering
  \includegraphics[width=0.99\textwidth]{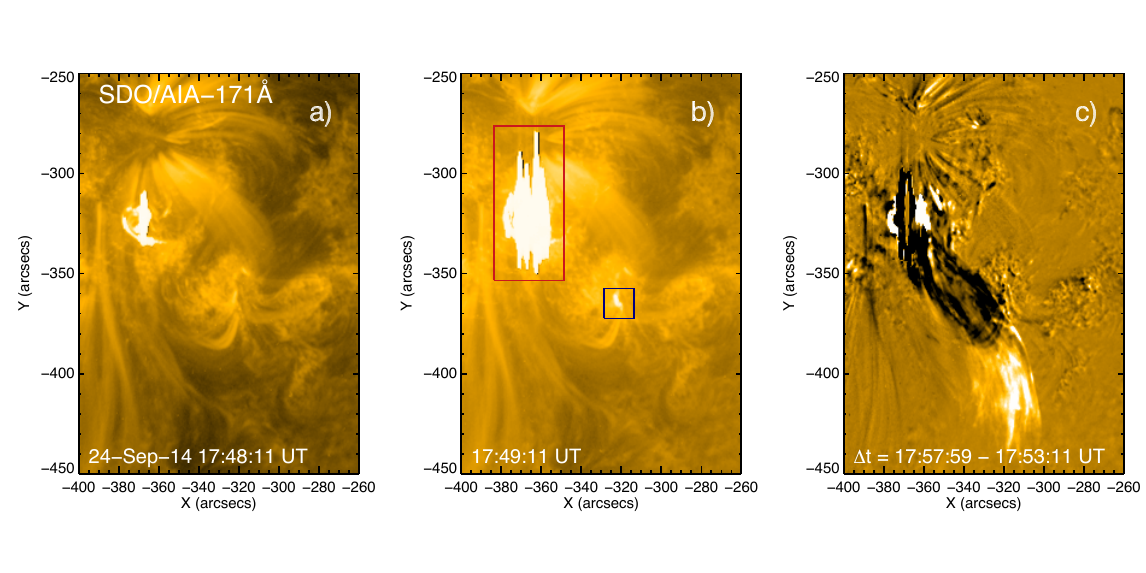}
        \caption{Temporal evolution of the spine-fan flare observed at 171 \AA~by the SDO/AIA on 24 September 2014. a) The initial phase of the spine-fan flare (Fig.~\ref{fig:mcmp}) recorded at 17:48:11 UT. b) The main flare (red, larger box) and the remote brightening source (blue, smaller box). c) Running-difference image to highlight the presence of an eruptive plasma.
        \label{fig:aia171}
        }
\end{figure*}

Fig.~\ref{fig:aia171}(a-c) shows the evolution observed by SDO/AIA in the corona at 171 \AA. In panel~\ref{fig:aia171}a, at 17:48:11 UT, in addition to the bright regions highlighting the same strong footpoints observed in Fig.~\ref{fig:mcmp}, a circular ribbon is visible. Fig.~\ref{fig:aia171}b, taken at 17:49:11 UT, shows the presence the main flare source (in the red box) that saturated the 171 \AA~detector, and a remote brightening (indicated by a small box). The remote brightening is also detected in other EUV channels, such as 335\AA. In Fig.~\ref{fig:aia171}c, we show a difference image between a time during the gradual phase (17:53:11 UT) and the end of the flare (17:57:59 UT) to make the ejection material more clear: the darker regions show where the erupting material was before 17:57:59~UT, while the brighter regions shows this material at this instant in time.

It is usually suggested that circular ribbons appear due to the formation of a `spine-fan' magnetic configuration \citep[e.g.][]{Lau_1990_1990ApJ...350..672L, Priest_1996_1996RSPTA.354.2951P, Yang_2020} in which field passes through a three-dimensional coronal null point forming a dome-like `fan' separatrix surface that intersects the chromosphere, giving the circular ribbon, and an extended `spine' field extending from the null to a remote region. Many flares of this type have been observed \citep[e.g.][]{2001ApJ...554..451F,2009ApJ...700..559M, wang_2012_2012ApJ...760..101W,Deng_2013_2013ApJ...769..112D, Yang_2020}, and the null point, or its vicinity, is often invoked as a likely electron acceleration region.

\begin{figure}
\centering
  \includegraphics[width=0.45\textwidth]{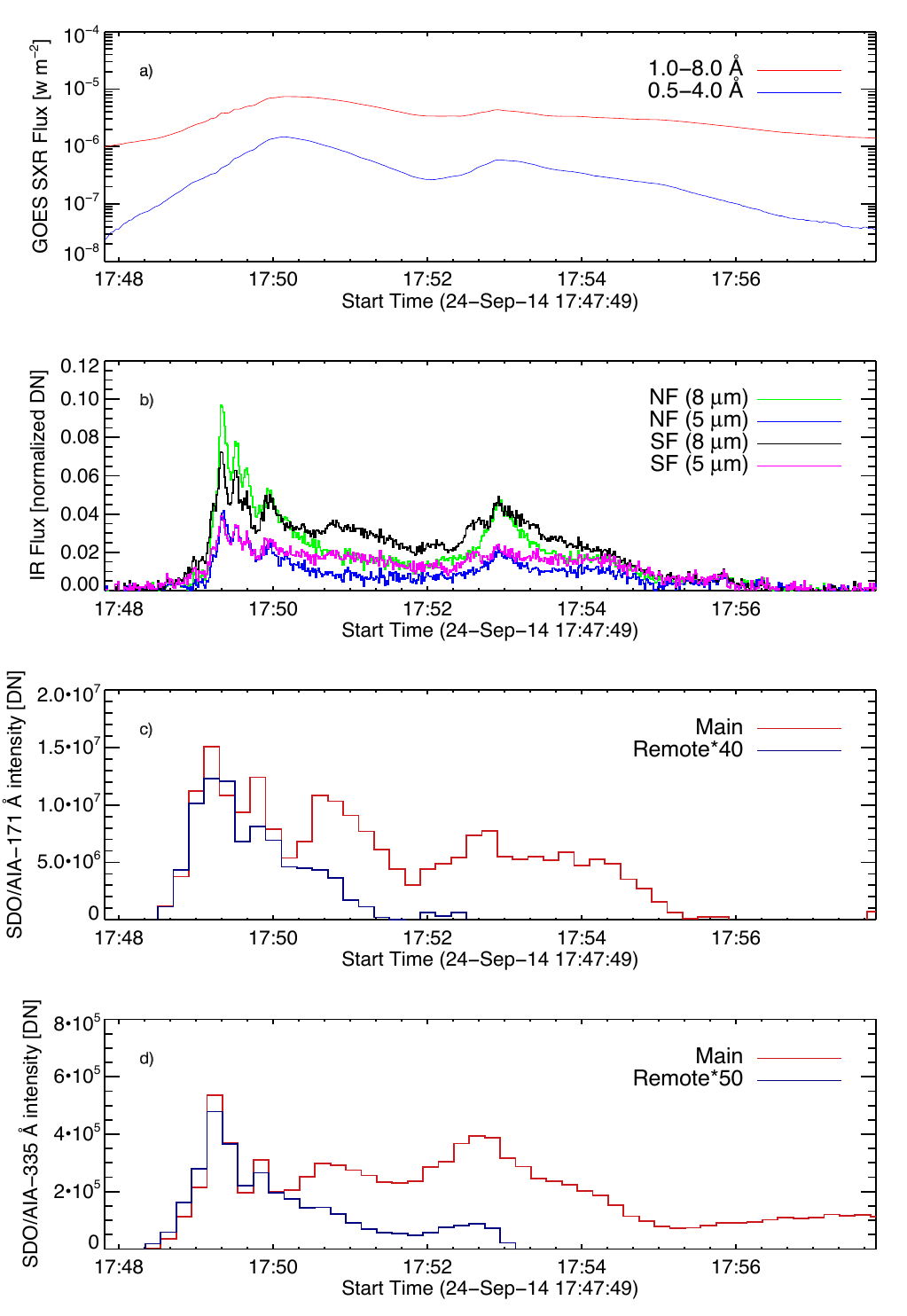} 
        \caption{Temporal profile at SXR (0.5-4.0 and 1-8~\AA), IR (8.2 and 5.2~\AA), and EUV (171 and 335~\AA) of the solar flare occurred on 24 September 2014 observed by GOES satellite, McMath-Pierce telescope, and SDO/AIA. The light curve excess at 171 and 335~\AA\ of the remote brightening are multiplied by 40 and 50, respectively, for better visualization.
        \label{fig:lc}
        }
\end{figure}

In Fig.~\ref{fig:lc} we compare the lightcurves of the various emissions between 17:47:49 - 17:57:00~UT: (a) Soft X-rays (SXR, 1-8 and 0.5-4 \AA) from the \emph{Geostationary Operational Environmental Satellite} (GOES) X-ray sensor, (b) IR (8.2 and 5.2 $\mu$m) for both NF and SF sources, (c) SDO/AIA 171~\AA, and (d) SDO/AIA 335~\AA. Both the main and remote sources are included in panels (c) and (d). For the IR sources we computed the lightcurves considering the total emission of a rectangular area of $4\times4$ pixels around NF and SF over the main flare region, and the pre-flare emission was subtracted, so that each lightcurve is the flare excess. The IR curves are analyzed in more detail in Section~\ref{sec:IRdata}. 

For the EUV sources, the lightcurve were obtained by summing over the the main flare region (red, large box in Fig.~\ref{fig:aia171}), and the remote brightening (blue, small box in Fig.~\ref{fig:aia171}). The pre-flare intensity was also subtracted from these curves. The EUV time profiles indicate that the evolution of the main and remote sources are strongly associated, at least until approximately 17:50~UT. We note that a second flare occurs after this time, evidenced by the second peak in the SXR curves. Although a circular ribbon is present in this event due to a likely spine-fan magnetic configuration, the main energy deposition regions are marked by two HXR footpoints which are co-spatial with the IR and white-light sources, denoting the regions where most of the accelerated electrons collide with the chromospheric plasma  \citep{2016ApJ...819L..30P}. In the case of a spine-fan configuration, these footpoints should be a subset of locations where the fan field enters the chromosphere and thus must be connected to the null within the magnetic fan, which is often though of as a favourable location for particle acceleration.
Therefore, they respond on the timescales relevant to the energy release and transport processes in this geometry. If we assume that these two footpoints, NF and SF, observed in IR and HXR (and the brightest regions in the EUV/UV circular ribbon) are connected to a null where particle acceleration takes place then they should also exhibit correlated behaviour.

\begin{figure*}
\centering
  \includegraphics[width=\textwidth]{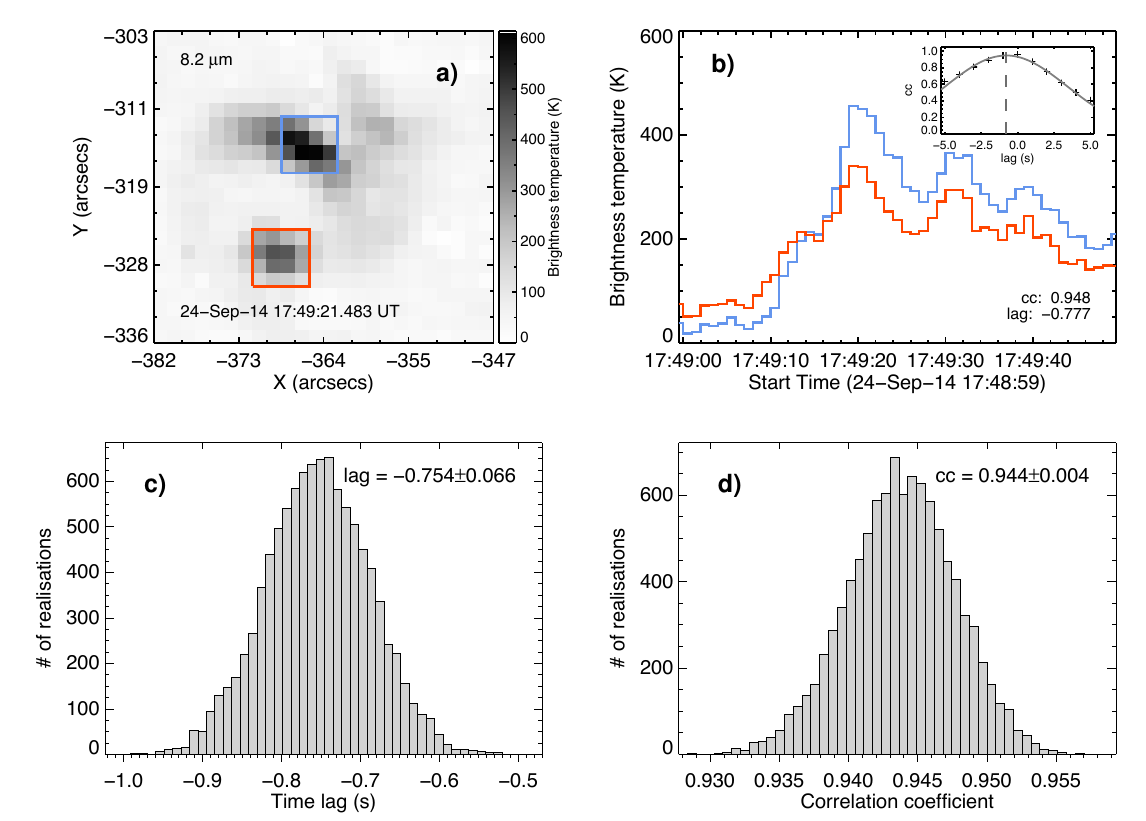}
        \caption{Time-delay analysis of IR 8.2 $\mu$m. (a) IR 8.2 $\mu$m image overlaid with the integration boxes for the two footpoint sources. (b) The resulting light curves from the northern (blue) and southern (red) footpoints, with the sub-panel showing the result of the lagged cross-correlation. The lower panels illustrate the Monte Carlo method for generating uncertainties in the (c) lag and (d) correlation coefficient values.
        \label{fig:timing8}
        }
\end{figure*}

\begin{figure*}
\centering
  \includegraphics[width=\textwidth]{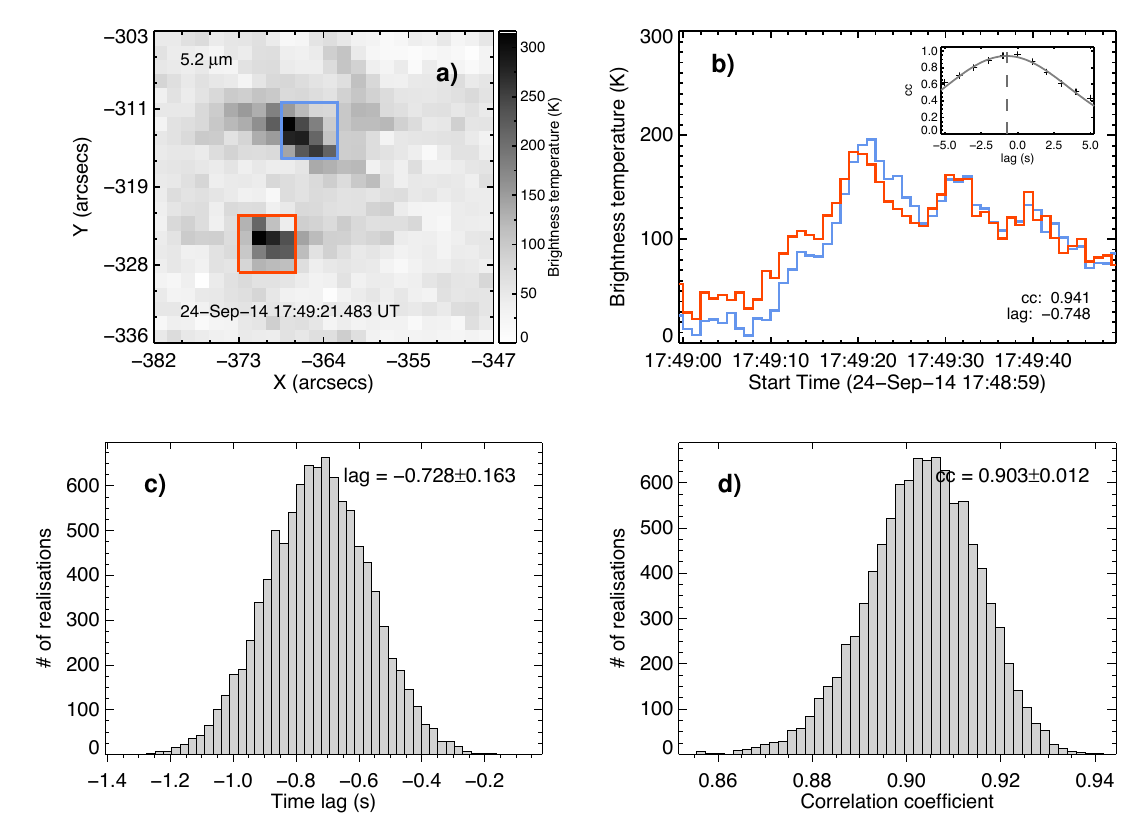}
        \caption{Same as Figure~\ref{fig:timing8} but for IR 5.2 $\mu$m.
        \label{fig:timing5}
        }
\end{figure*}

\ifpaper
\subsection{Time delays of mid-IR data}
\label{sec:IRdata}

We have used datacubes consisting of 1-s sampling of two broad mid-infrared continuum bands, centered at 5.2 $\mu$m and 8.2 $\mu$m, taken with Quantum Well Infrared Photodetectors (QWIP) in a developmental setup at an auxiliary feed telescope of 0.76~m diameter at the National Solar Observatory's McMath-Pierce telescope, with re-imaging to a plate scale of 0.76$''$/pixel in a 320$\times$256 array. As pointed out in \cite{2016ApJ...819L..30P}, the infrared observations do not have absolute coordinate references, so the images were empirically co-aligned to images from SDO/AIA images using the sunspot umbrae as references. Also, the absolute timing of the IR data is uncertain due to uncertain clock synchronisation; however, the relative timing information between the two IR channels is precise.

The time-series photometry in our analysis uses boxes of 4$\times$4 pixels around each of the two main IR sources, as indicated in blue and red for the northern and southern sources in Fig.~\ref{fig:timing8} and~\ref{fig:timing5}, for 8.2 and 5.2 $\mu$m respectively. For each box, we obtained light curves for the flare excess by averaging the pixel values and subtracting the pre-flare emission, shown in the Fig.~\ref{fig:timing8}b and ~\ref{fig:timing5}b. 

For each wavelength, we cross-correlated the light curves of the two sources during the impulsive phase (between 17:48:59 and 17:51:59~UT) using the IDL function \verb!c_correlate!, which returns the Pearson linear correlation coefficient as a function of the lag. To obtain lag values finer than the 1-s cadence of the data, we found the peak correlation by fitting the correlation coefficient in the range $\pm$5 s of lag with a Gaussian \cite[e.g. ][]{Lai753016,Zhang2006,Qin4697676,Dabrowski2009A&A...504..565D,Rosseel9610902}. Similar sub-resolution methods for time delay estimation rely on interpolating the lag correlation values instead of fitting a function \citep[e.g.][]{1995ApJ...447..923A,Aschwanden1996ApJ...468..398A,Altyntsev2019ApJ...883...38A}. This is shown in the subpanels of the Fig.~\ref{fig:timing8}b and ~\ref{fig:timing5}b, with the crosses indicating the output from the cross-correlation and the solid curve being the best-fit Gaussian function. This gives correlation coefficients of 0.94 and 0.95, with lags of -0.75~s at 5.2~$\mu$m, and -0.78~s at 8.2~$\mu$m, respectively. The negative sign indicates that the northern (blue) source lags behind the southern (red) source.

We estimate uncertainties in the measured time delays due to the effects of data noise using a Monte Carlo analysis. Taking a non-flaring period between 17:57:49 and 18:01:08~UT, we construct the histogram and the cumulative distribution function (CDF) of the time series to obtain its noise properties. From this distribution we randomly draw positive and negative noise values to add to the original time series. Multiple time-series realisations are then cross-correlated; histograms of the lag and correlation coefficient for 10,000 realisations are shown in Fig.~\ref{fig:timing8}c and \ref{fig:timing8}d
 for 8.2~$\mu$m, and Fig. \ref{fig:timing5}c and \ref{fig:timing5}d for 5.2~$\mu$m. Taking the mean and standard deviation of these distributions gives $-0.754 \pm 0.066$~s for the lag at 8.2~$\mu$m, with a correlation coefficient of $0.944 \pm 0.004$. For 5.2~$\mu$m, the values for the lag and correlation coefficient are $-0.727 \pm 0.163$~s and $0.903 \pm 0.012$ respectively. The estimated uncertainties for 5.2~$\mu$m are larger because of the slightly lower signal-to-noise ratio of the data, when compared to the 8.2~$\mu$m data.
 
 \fi
 

\begin{figure*}
    \centering
    \includegraphics[width=\textwidth]{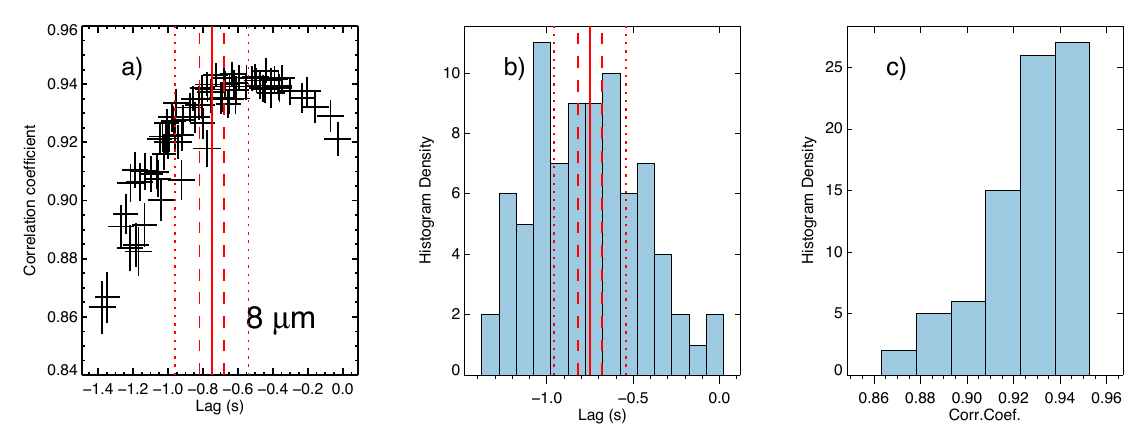}
    \includegraphics[width=\textwidth]{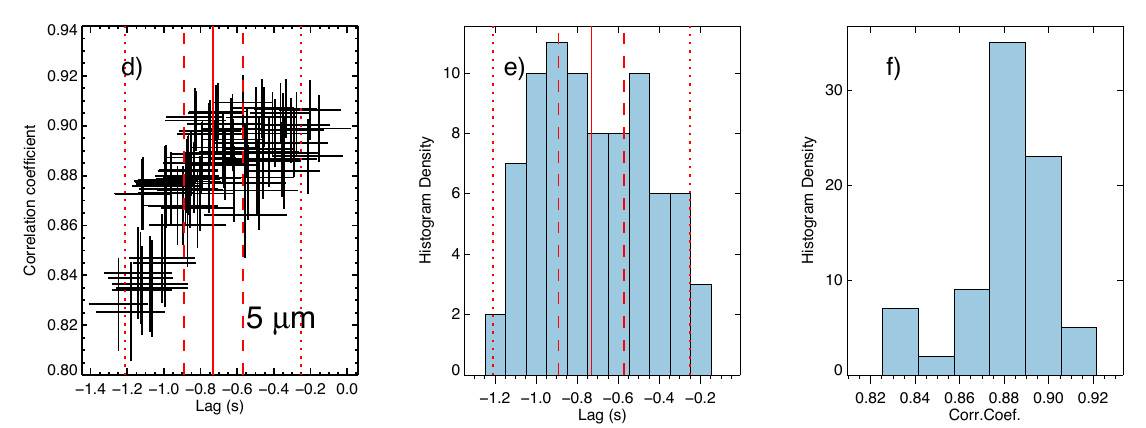}
    \caption{Results of testing the combination of shifting one pixel of each selection box (up and down, left and right) in the IR images, following the same Monte Carlo procedure to estimate uncertainties. (a) correlation coefficient versus the time lag for 8.2~$\mu$m; histograms of the (b) time lag and (c) correlation coefficient. The same results for 5.2~$\mu$m are shown in panels d, e, and f. The vertical red lines indicate the time lag found for each wavelength, along with its 1 $\sigma$ (dashed) and 3 $\sigma$ (dotted) uncertainties.}
    \label{fig:boxshift}
\end{figure*}
 
Lastly, to verify the influence of the selection boxes around each IR footpoint, we reevaluated the time lags for each wavelength channel shifting one pixel in each box, in one direction at a time (up, down, left, right), for a total of 81 new realizations of the time lag analysis. The results are shown in Figure~\ref{fig:boxshift}: correlation coefficient versus the time lag for 8.2~$\mu$m and 5.2~$\mu$m; histograms of the time lag and correlation coefficient. We find a wider spread of time lag values, from 0 to 1.4 seconds, but statistically more concentrated around our initial results for the time lag found for each wavelength, indicated by the vertical red lines. Also shown are the 1-$\sigma$ (dashed) and 3-$\sigma$ (dotted) uncertainties on these time lags.

\ifpaper
\subsection{Hard X-rays} \label{sec:hxr}
Since its launch in 2002, RHESSI \citep{LinDennisHurford:2002} has observed the Sun in X-rays from 3~keV up to 17~MeV, providing images and spectra of solar flares. RHESSI is equipped with an attenuator to avoid pulse pileup during times of high flux of incident low-energy photons (the few-keV thermal spectrum).
Here we use RHESSI data to obtain information about the accelerated electrons during the flare. We now describe the methodology to use HXR images and spectra to obtain the characteristics of the accelerated electrons in both northern and southern flare sources. We created HXR images with the CLEAN algorithm \citep{HurfordSchmahlSchwartz:2002} using data from RHESSI's front detectors 1 to 8, setting the \verb!clean_beam_width_factor! to 1.5, with an integration time of 28 seconds, in the time interval 17:49:05 to 17:49:33 UT, that covers the impulsive phase of the event, at the energy bands 12--15, 15--20, 20--25, 25--30, 40--50, and 50--70 keV. During this time interval, the thin attenuator (state A1) was active. 

The image datacube was loaded into the standard spectral analysis software OSPEX \citep{SchwartzCsillaghyTolbert:2002}, part of the SolarSoftware \citep[SSW;][]{FreelandHandy:1998} in IDL. Within OSPEX, we defined a rectangular region for the HXR northern and southern sources as shown in Fig. \ref{fig:spex}a, which are co-spatial with the IR sources. For each box, the photon intensity is summed, for each energy band, resulting in the photon spectrum for the selected source. The spectra were then fitted with a single power-law model $I(\epsilon)=K\epsilon^{-\gamma}$, as shown in Fig. \ref{fig:spex}b. The parameters for the best fit are $K_N=0.55 \pm 0.04$ photons s$^{-1}$ cm$^{-2}$ keV$^{-1}$ at 50~keV, $\gamma=1.8 \pm 0.1$, and $K_S=0.35 \pm 0.04$ photons s$^{-1}$ cm$^{-2}$ keV$^{-1}$ at 50~keV, $\gamma=2.1 \pm 0.2$, for the northern and southern sources respectively. 

Assuming the thick-target model \citep{1971SoPh...18..489B,1972SoPh...24..414H}, a power-law photon spectrum with a spectral index $\gamma$ is produced by an electron distribution, also with a power-law form, with a spectral index $\delta$, and they are related according to $\delta=\gamma+1$. Therefore, the estimated average spectral index of the electron distribution for each of the two sources are $\delta_N=2.83\pm0.13$ and $\delta_S=3.11\pm0.15$.

We have also analysed the spatially integrated X-ray spectrum (Fig. \ref{fig:spex}c), integrated for 20 seconds during the impulsive phase 17:49:08 to 17:49:28 UT. The spectrum was fitted with an isothermal component plus a double power-law thick-target model. The isothermal model assumes that the thermal flare plasma, with a density $n$ and temperature $T$ and occupying a volume $V$, emits X-rays by brems\-strahlung. The contribution from spectral lines for a solar coronal abundance is also computed, from the CHIANTI package \citep{LandiYoungDere:2013}, also part of SSW. The non-thermal model takes the form of a photon spectrum produced by an electron distribution $F(E)$ in the form of a broken power-law undergoing bremsstrahlung in a collisionally thick target \citep{1971SoPh...18..489B,1972SoPh...24..414H}. The parametric $F(E)$ is defined by the spectral indices $\delta_L$ and $\delta_H$ below and above an energy break $E_\mathrm{brk}$ respectively. The model is normalized by the total rate of electrons $N$ above a reference energy $E_c$, which indicates the lowest electron energy necessary to explain the observed spectrum. The best fit parameters found are: emission measure $\xi_\nu=n^2V=3.4 \pm 0.9 \times 10^{47}$ cm$^{-3}$ at $T=17.5 \pm 1.1$ MK for the isothermal model, and $N=4.6 \pm 2.1 \times 10^{33}$ electrons s$^{-1}$, $\delta_L=2.4 \pm 0.4$, $\delta_H=3.8 \pm 0.1$, $E_\mathrm{brk}=83 \pm 15$ keV, $E_c=15 \pm 30$ keV for the non-thermal model. The total power contained in the non-thermal electron distribution is found by $P=\int_{E_c}^\infty E F(E) dE$, giving $P\approx 2.6 \times 10^{27}$ erg s$^{-1}$. 

For completeness, we have also verified the evolution of the HXR emission from each source, producing light curves, shown in Fig. \ref{fig:spex}d, for the square regions defined in Fig. \ref{fig:spex}a and a time sequence of RHESSI images reconstructed at 20--70 keV using CLEAN. As expected, for the 4-second cadence of these images, no measurable time delays are observed between the two time-series.
\begin{figure*}
\centering
  \includegraphics[width=0.45\textwidth]{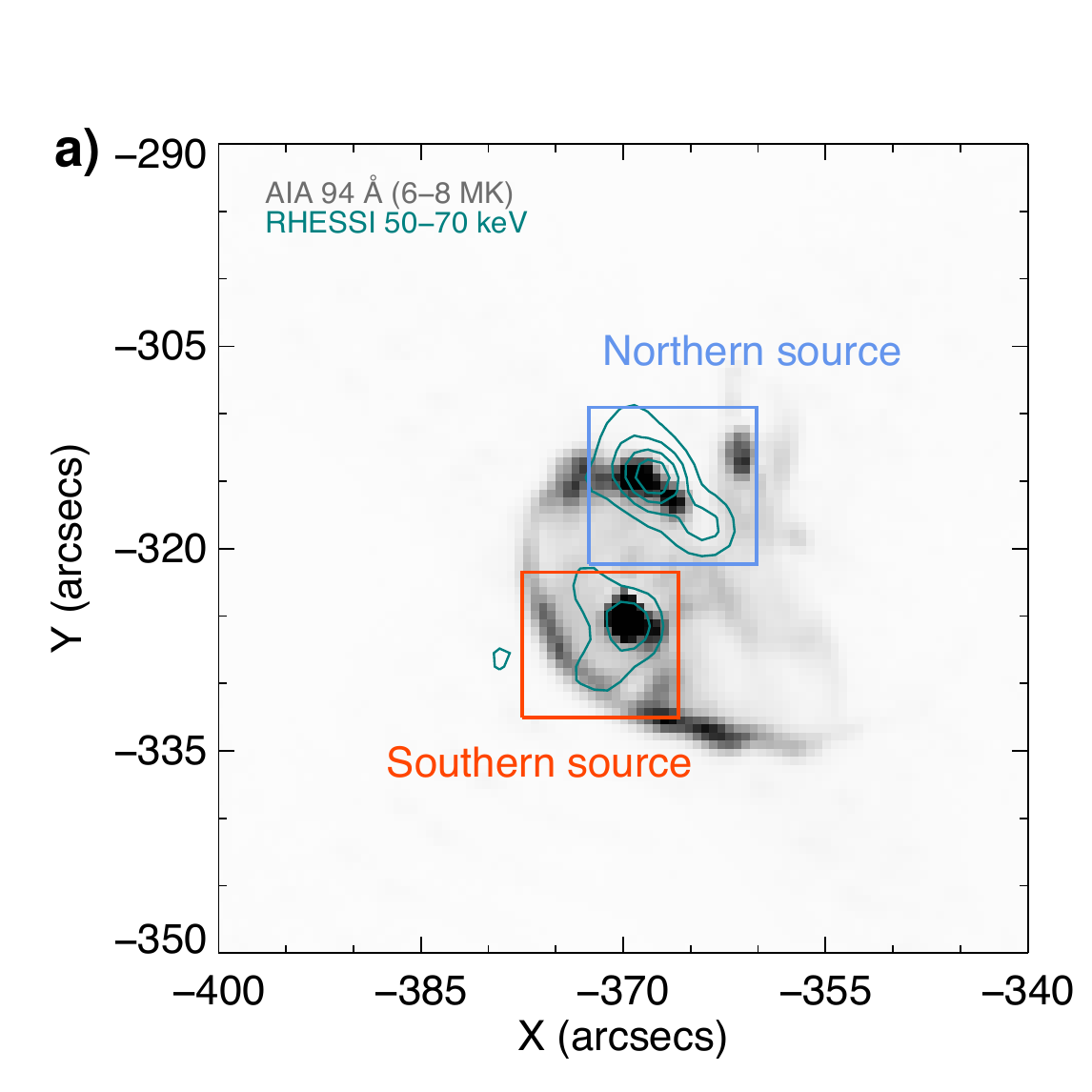}
  \includegraphics[width=0.45\textwidth]{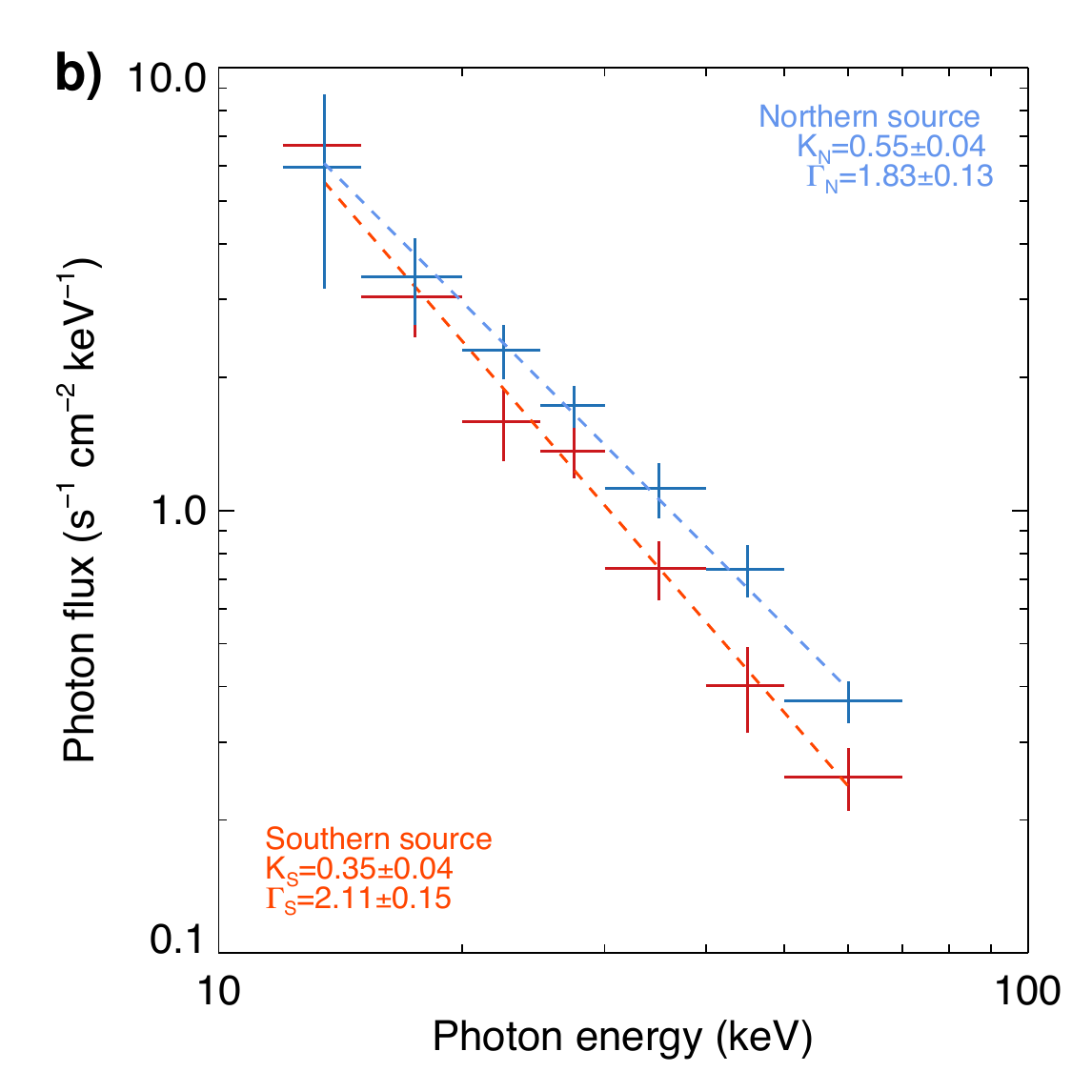}
  \includegraphics[width=0.45\textwidth]{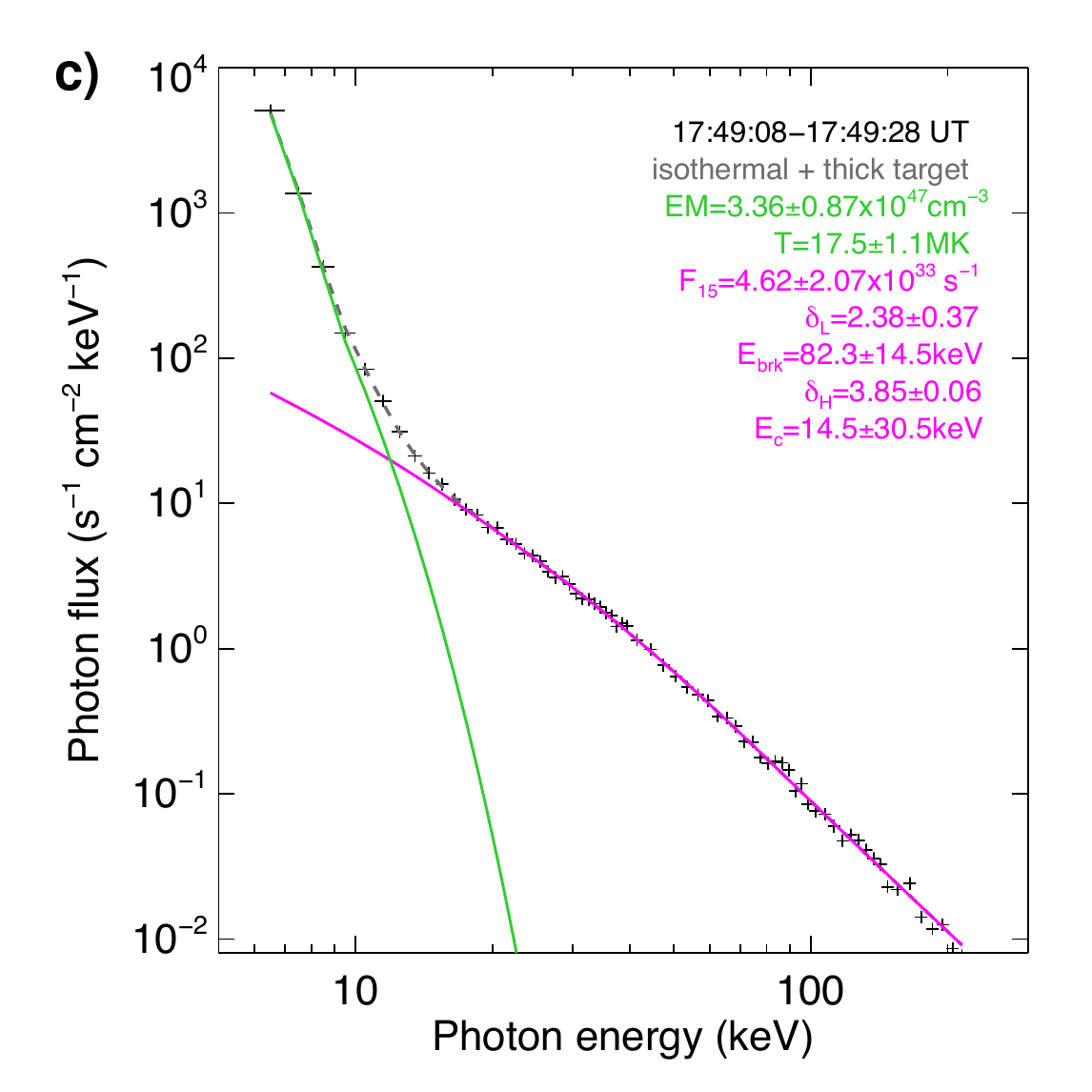}
    \includegraphics[width=0.45\textwidth]{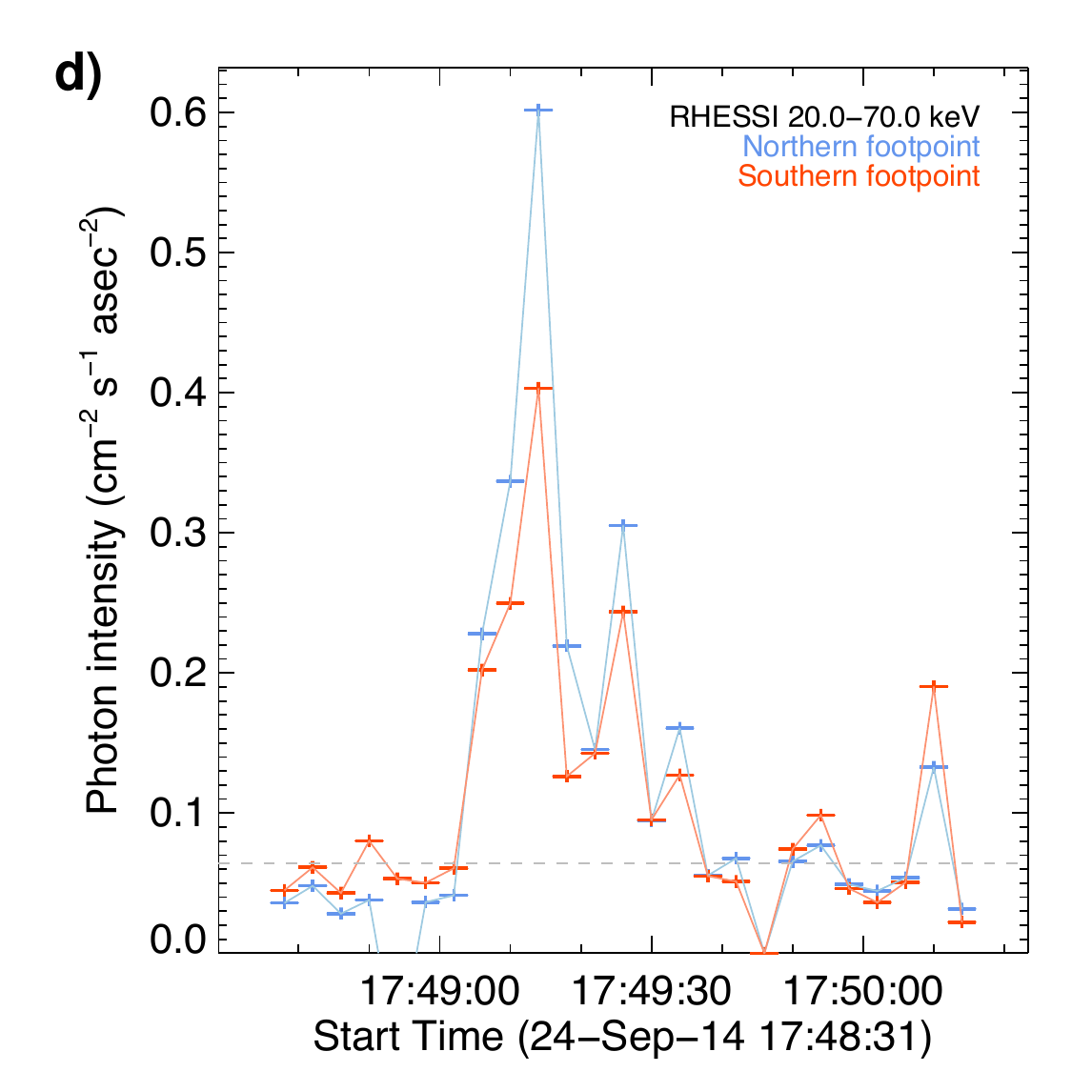}
        \caption{{Analysis of the HXR emission of the footpoint sources.} (a) HXR sources at 50-70 keV (contours) over AIA 94~\AA~image (inverse colours). (b) RHESSI spatially-integrated HXR spectrum, fitted with isothermal (green) and non-thermal (magenta) components. (c) HXR spectrum of the HXR northern (blue) and southern (red) sources, fitted with a power-law. (d) RHESSI HXR 20--70 keV 4-seconds curves of each footpoint source.
        \label{fig:spex}
        }
\end{figure*} 
\fi 

\subsection{Possible implication of the observed IR time lags.} 
In the electron-beam model, a time lag between footpoints corresponds to a difference in travel time for electrons traveling along the magnetic field from the acceleration site. In the simple case of a symmetrical loop with acceleration at its mid-point there should be zero lag between the footpoints, while an offset acceleration site (that is, the acceleration site is located some distance from the midpoint, along one of the legs) should have a finite lag. 
Using the better-determined lag of $0.75$~s at $8.2\ \mu$m, and an estimated speed corresponding to 30~keV electrons of $v\sim 10^{10}\ \rm{cm~s}^{-1}$ gives a difference in the distances traveled of $\Delta l$ = 75,000~km if the electrons are free-streaming. This is much larger than the footpoint separation of 8,700~km ($\simeq$12$''$). 
The IR time lags thus appear to be inconsistent with those achievable assuming electrons accelerated into a single loop joining the two IR footpoints. However there are several potential ways in which a time lag could be introduced, which must be estimated. These are (i) the dynamics of the electrons in the corona, including their spiralling motion and magnetic trapping; (ii) the formation of the observed IR emission and (iii) the geometry of the field connecting the footpoints to each other and to the accelerator. These will be examined in the following three Sections. 

\section{Magnetic Field Geometry}~\label{sec:geom}
\begin{figure}[!h]
\centering
  \includegraphics[width=0.45\textwidth]{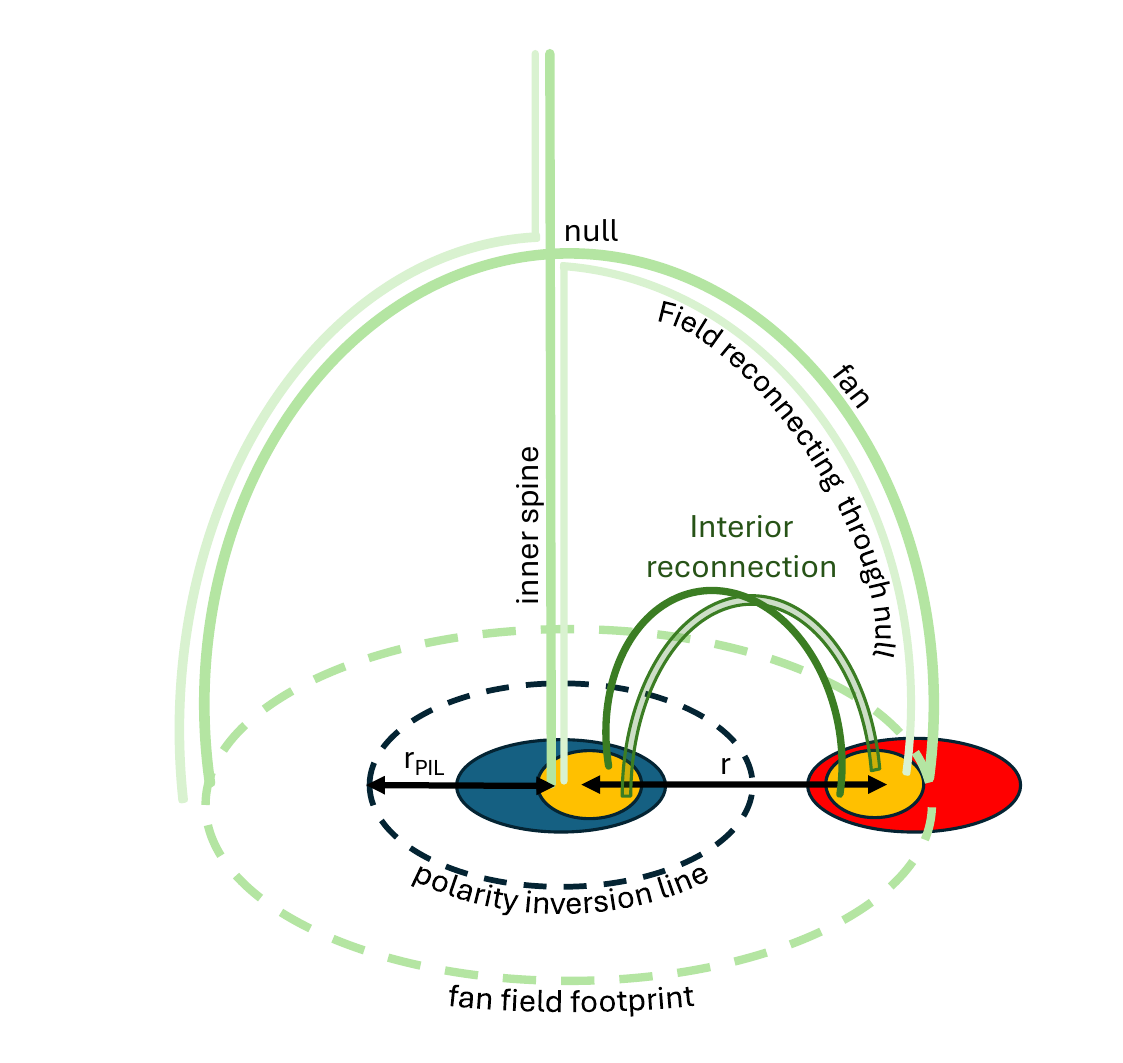}
        \caption{A sketch of two possible locations for reconnection in a circular ribbon topology. Blue and red indicate the two field sources, and yellow patches are the IR/HXR footpoints. Example field lines are shown in different shades of solid green lines, along with the photospheric polarity inversion line (black dashed) and the photospheric footprint of the fan field (green dashed). Reconnection could happen interior to the separatrix dome formed by the fan field, or through the null.  
        \label{fig:topo}
        }
\end{figure}
The strong timing correlation between the two main IR footpoints shown in panels b) in Figures~\ref{fig:timing8} and ~\ref{fig:timing5} provides clear evidence that they are linked, presumably by the magnetic field. The length of the connecting field, and the magnetic field convergence, are both important in understanding the timing lag; the latter because it influences particle trapping in the corona. As noted, the event demonstrates a circular ribbon consistent with a spine-fan magnetic topology. One of the two IR sources is coincident with negative field and the other with part of the surrounding positive field, as shown observationally in Figure~\ref{fig:hmi1} and sketched in Figure~\ref{fig:topo}. A roughly circular polarity inversion line (PIL) extends around most of the southern source in the negative field. The observations are consistent with the southern IR/HXR source being close to or coincident with one end of the spine field, and the northern one being close to or coincident with the intersection of part of the fan separatrix surface with the photosphere.  A similar configuration was studied with field extrapolations, magneto-hydrodynamic (MHD) simulations and UV/HXR and magnetogram observations by \cite{2009ApJ...700..559M} and \cite{2012A&A...547A..52R}. The latter authors proposed that their strong HXR/UV sources, situated in equivalent locations to our strong HXR/IR/UV sources, were the result of reconnection at a quasi-separatrix layer (QSL) within the separatrix dome \citep[a possibility also discussed by][]{2001ApJ...554..451F}. Reconnection at the null, and in the QSLs surrounding both it and the spine fields, was identified by \cite{2009ApJ...700..559M} and \cite{2012A&A...547A..52R} as responsible for the main circular ribbon and the remote source appearing later in the event, but did not appear to provide a natural explanation for the strong  HXR/UV sources. 
 
In the case of (i) reconnection at a QSL interior to the separatrix dome, a reasonable lower limit to the length of the field connecting sources is a semicircular loop of length $l = \pi r/2 = 13,700$~km where $r =8,700$~km is the separation of the sources. However, in principle the loop could extend higher, towards the top of the separatrix dome. A hemispherical separatrix dome has been assumed in some studies \citep[e.g.][]{2022ApJS..260...19Z} at least as a starting configuration, but \cite{2009ApJ...691...61P} demonstrate in simulations that if the field is stressed by twisting the separatrix dome inflates and the null can rise to a pre-eruption height of around 5 times the radius of the PIL ($r_\mathrm{PIL}$) from the central field source. Assuming a half-ellipsoid shape for the separatrix dome, with semi-minor axis $r_\mathrm{PIL}$ and semi-major axis $5 r_\mathrm{PIL}$ gives the outer perimeter of the dome as $10.5 r_\mathrm{PIL}$ (using the Wolfram|Alpha online calculator). If we further assume that $r_\mathrm{PIL} = r/2$, i.e. the PIL lies midway between the two IR sources, then the outer perimeter length is $5.25r = 42,000$~km. The maximum length for a loop reconnecting at an interior QSL might be around half of this.

The outer perimeter length also gives an estimate for the length of the linking field in the case of (ii) reconnection through the null. Arguably the maximum length of the reconnecting field would be less than this, as part of it would follow the inner spine, but the estimate allows for some winding of the field, and asymmetry in the spine-fan configuration. In the case of (i) interior reconnection, identified by previous authors as the explanation for their HXR sources the full loop length corresponds to a time-of-flight from one footpoint to the other - i.e. a maximum lag - for 30\,keV electrons of 0.14\,s; in the case of (ii) null-point reconnection it corresponds to 0.42\,s. Both are substantially shorter than the observed lag of 0.75\,s. If electrons are accelerated closer to the midpoint of the connecting field in case (i), or at the null in case (ii), the possible lag of course reduces.

\section{Electron transport simulations}\label{sec:transport}

Firstly, spiraling motion of magnetized electrons allows larger delays for electrons with larger pitch angles (angle of velocity vector to the magnetic field) which in effect traverse a longer path for a given physical distance along the loop. If coronal propagation is approximately collisionless, and assuming a semi-circular loop joining the footpoints, with a length $l \approx 13,700$~km, the observed delay could be explained by electrons traveling with zero pitch angle along one leg from the loop's midpoint, but with pitch angle $\theta \approx 85^\circ $ along the other. However, the field converges with decreasing height above the surface (a `magnetic bottle') so particles with such a high pitch angle would be trapped in the corona and unable to reach and excite the chromosphere -- unless the ratio of field strengths at the IR footpoint and the coronal acceleration site is between 1 and $1/\sin^2\theta = 1.01$.

Delays can also happen due to electrons being magnetically trapped in the corona and leaking out slowly due to collisional or non-collisional scattering. The trap lifetime depends on the scattering regime, which is determined by the pitch-angle diffusion and the geometric properties of the trap \citep{1997A&A...326.1259F}. 
\ifpaper
We employed Fokker-Planck simulations \citep{HamiltonLuPetrosian:1990} to evaluate the timing and relative number of electrons at each footpoint. The simulations calculate the evolution of an electron distribution function $f(E,\mu,s,t)$ as function of energy $E$, cosine of pitch-angle $\cos \theta=\mu$, position $s$ and time $t$ considering collisional energy losses and pitch-angle scattering $D_{\mu\mu}$, and magnetic mirroring, as described in Equation \ref{eq:fokkerplanck}.
\begin{equation}
\frac{\partial f}{\partial t}=-\mu c \beta \frac{\partial f}{\partial s}-\frac{\partial}{\partial \mu} \dot{\mu}f-\frac{\partial}{\partial E}\dot{E}f+\frac{\partial}{\partial \mu}\left(D_{\mu\mu}\frac{\partial f}{\partial \mu}\right)+S(E,\mu,s,t),
\label{eq:fokkerplanck}
\end{equation}
where $\dot{E}=-4\pi ncr_0^2\ln \Lambda / \beta$ is the energy loss term by Coulomb collisions, $\dot{\mu}=-\frac{1}{2}\beta c (1-\mu^2)(\mathrm{d} \ln B/\mathrm{d}s)$ is the rate of change of $\mu$ due to magnetic mirroring, $D_{\mu\mu}=4 \pi n c r_0^2 \ln \Lambda (1-\mu^2)/\beta^3 \gamma^2$ is the pitch-angle scattering coefficient, with thermal plasma density $n$, speed of light $c$, $\beta=v/c$, Lorentz factor $\gamma$, the electron classical radius $r_0$, and Coulomb logarithm $\ln \Lambda$.

The numerical experiment was designed as follows: a loop was defined with full length $L=13,700$~km, and with an asymmetric magnetic field at the footpoints, $B_N$ and $B_S$. We assumed a magnetic field at the loop top $B_0=100 $~G, and uniform plasma density $10^9$~cm$^{-3}$.  We estimated the magnetic field values at the footpoints using the median values inside the boxed regions of the magnetogram from \textit{Helioseismic and Magnetic Imager} \citep[HMI,][]{2012SoPh..275..207S}, on board of SDO, as shown in Figure \ref{fig:hmi1}, where we find the values $B_N=110 $~G and $B_S=-400 $~G. 
\begin{figure*}
\centering
  \includegraphics[width=0.7\textwidth]{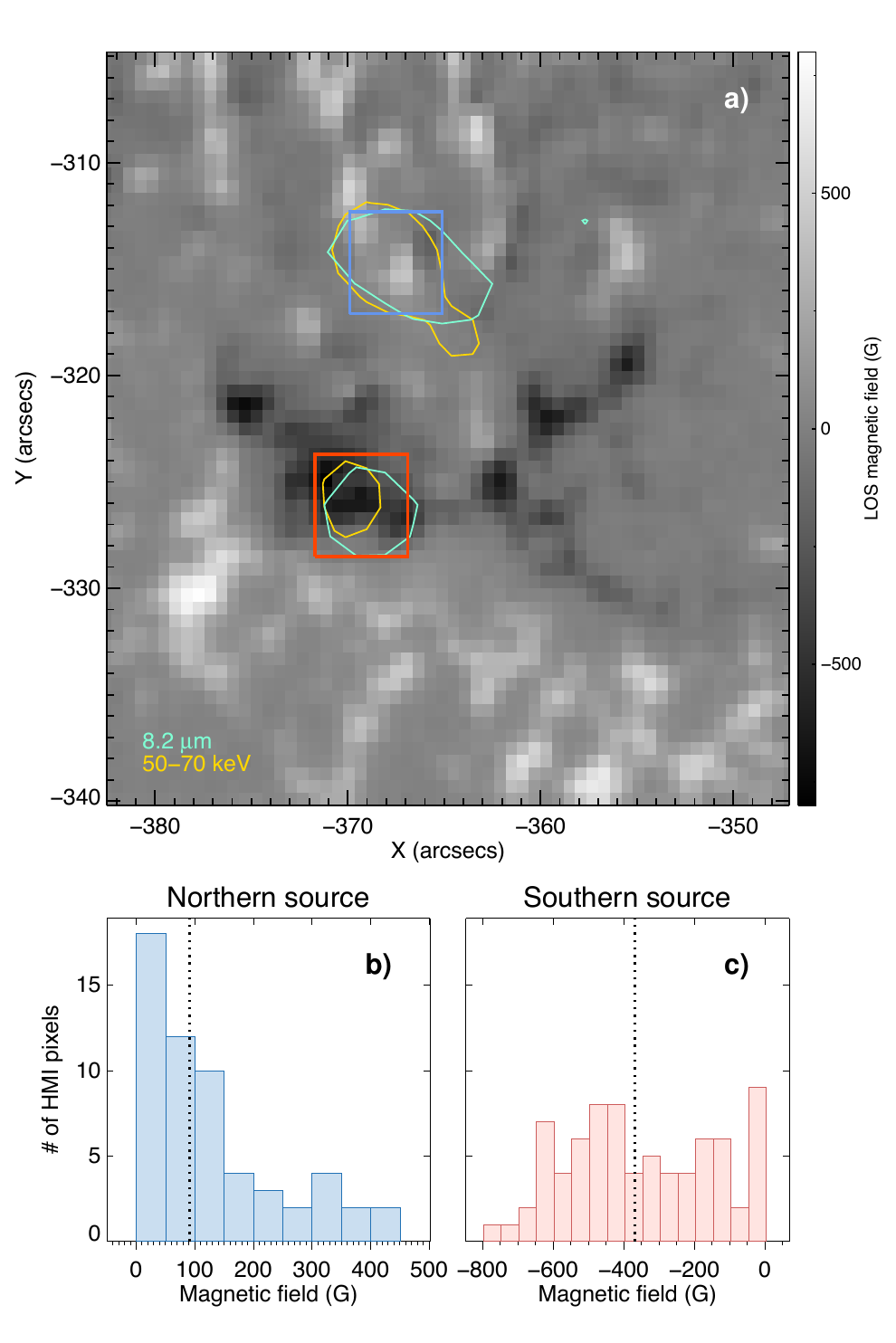}
        \caption{{Magnetic field configuration of the flaring region.} (a) SDO/HMI magnetogram of the flare region overlaid by 50\% contour levels of RHESSI 50--70 keV (gold) and IR $8.2~\mu$m (cyan). The blue and red boxes indicate the regions used to estimate the magnetic field. (b) Histogram of the photospheric magnetic field values inside the northern source box (blue) in panel (a). (c) Same as panel (b), for the southern source (red). The vertical dotted line indicate the median of the distribution, used in the Fokker-Planck simulations to set the field strength in the footpoints.
        \label{fig:hmi1}
        }
\end{figure*} 

The injection function $S(E,\mu,s,t)$ is described as a power-law distribution of electrons with $\delta=4$ injected in the corona, midway between the two main footpoints with a Gaussian time profile with a maximum at $t=1$~s and width $\tau=0.25$~s. Two pitch-angle distributions were considered: {\em isotropic} ($df(\mu)/d\mu=0$) and {\em field-aligned beam}, defined by Gaussians centred at $\mu=\pm1$ with width $\Delta \mu=0.3$. The time resolution of the simulation is $0.05$~s. The evolution of the pitch-angle distribution of 30 keV electrons at the looptop and each footpoint for the two beam models are shown in Fig.~\ref{fig:fokker1}. The evolution of electrons at other energies is very similar, as they travel almost collisionlessly through the coronal plasma in the magnetic loop. As expected, a larger number of electrons precipitate through the footpoint with weaker magnetic field, with a fraction only about 0.1 and 0.3 (for the isotropic and aligned-beam cases, respectively) of the electrons precipitating through the footpoint with the stronger magnetic field. 

We then calculated the thick-target 30~keV HXR photon flux from the electron distribution functions $f(E,\mu)$ at each footpoint using the angle-averaged bremsstrahlung cross-section \citep{1971SoPh...18..489B}. The resulting light curves for each footpoint were cross-correlated and shown in Fig. \ref{fig:fokker2}. The isotropic model in an asymmetric trap produces a time delay of 0.175~s in the HXR flux produced at the footpoints, while the aligned beam has a smaller time delay of 0.100~s in the simulation. The time delays are much smaller than those observed in the mid-IR data. Also, the ratio of the HXR emission is too high in the simulations, approximately 4 and 12 for the aligned-beam and isotropic cases, respectively, and thus unable to explain the observed HXR sources with similar intensities at 30--40~keV, with a ratio of $1.5\pm 0.3$. Increasing the trap asymmetry will increase the lag, but will also increase the relative electron numbers, and hence HXR flux asymmetry. That is, increasing the time-lag in the simulations via magnetic mirroring drives the ratio of the HXR flux from each source even further away from the observed value. Moreover, in the isotropic model the largest fraction of the electrons remains trapped close the to looptop, and so even in a relatively low density trap, thin-target coronal HXR emission would be present, contrary to the observations, where no HXR loop emission is observed during the impulsive phase.

Only the spatial variation of the magnetic field along the loop (i.e. convergence) is important for the magnetic mirroring. The HMI values for the magnetic field support the choice of loop asymmetry of $1:4$ between the two footpoints. This asymmetry could be higher (e.g. $1:8$), and that would produce bigger delays but the ratio of the number of electrons precipitating at each footpoint would also be much higher. On the other hand, a smaller $B$ asymmetry (e.g. $1:2$) would give a smaller electron number ratio, and hence a ratio of HXR at each footpoint closer to the observations, but the delays would be then too small, because most of the electrons would precipitate directly, as dictated by their time-of-flight only.
\begin{figure*}
\centering
  \includegraphics[width=\textwidth]{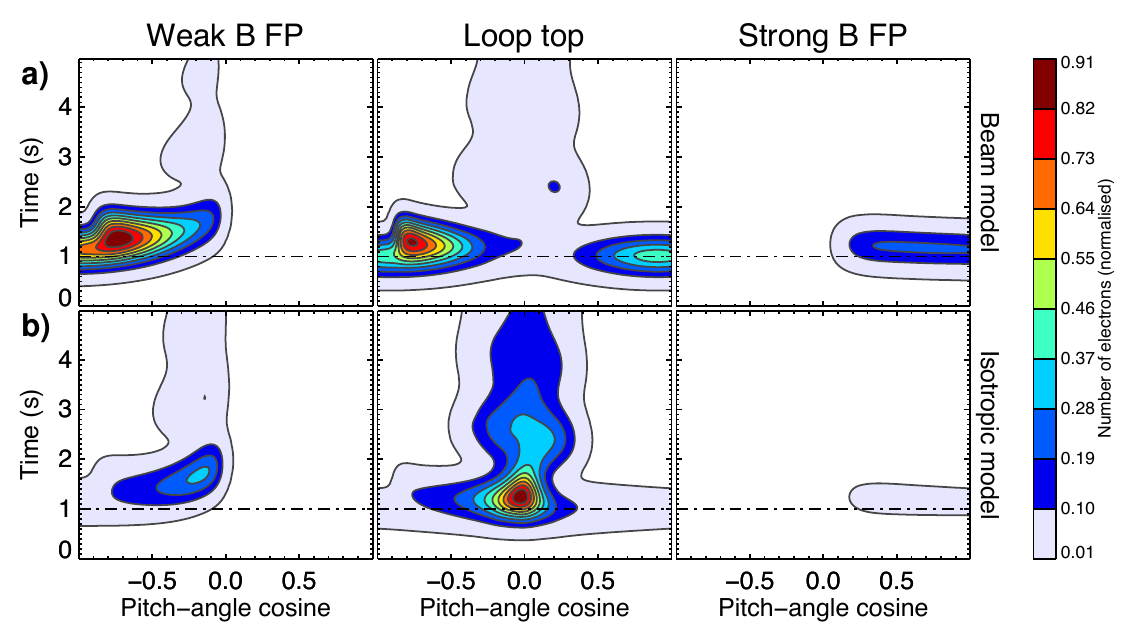}
        \caption{{Evolution of the pitch-angle distribution of 30~keV electrons at the coronal injection site} and at each footpoint (FP) for the two beam models (a) aligned beam model; (b) isotropic model. 
        \label{fig:fokker1}
        }
\end{figure*} 
\begin{figure*}
\centering
  \includegraphics[width=\textwidth]{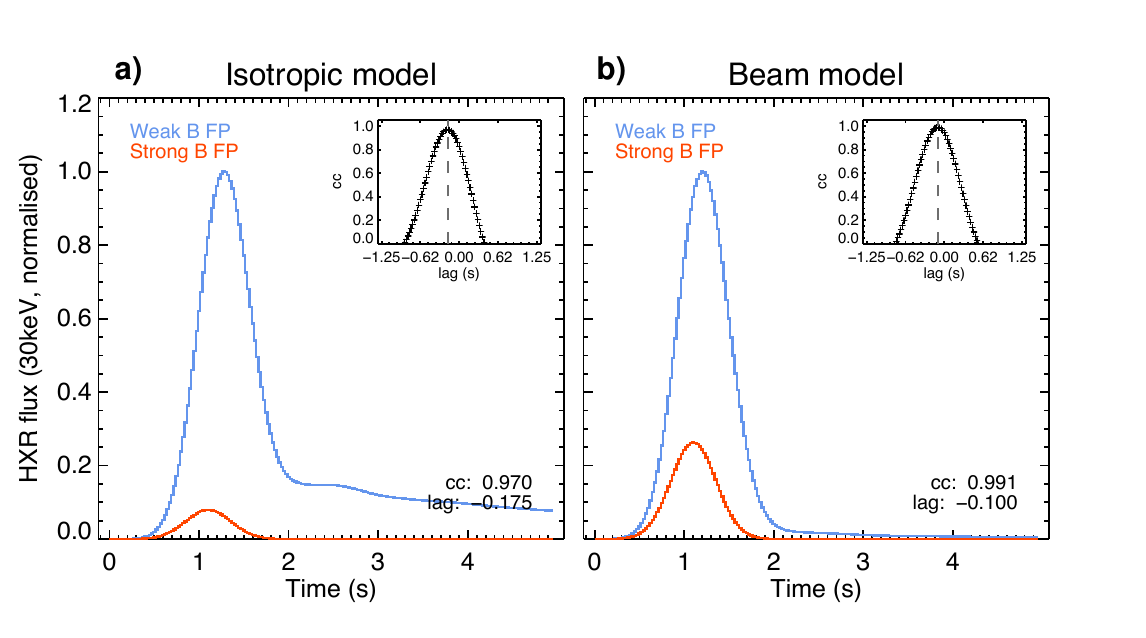}
        \caption{{Time evolution of the relative 30 keV HXR emission at the conjugate footpoints} (a) calculated from Fokker-Planck simulations for the isotropic injection. (b) Same as panel (a), but for the beam model. The sub-panels show the correlation coefficient as function of the lag.
        \label{fig:fokker2}
        }
\end{figure*} 
\fi

As expected, in the simulations, the electrons precipitate sooner at the weaker-field (northern) footpoint, in contradiction to IR observations in which the northern source lags the emission originating from the southern source. More electrons precipitate at the northern footpoint resulting in a larger HXR flux, which has the same sense as the observations but the simulated HXR intensity ratio at 30~keV is 4--12 (Fig.~\ref{fig:fokker2}), much higher than the observed ratio of $\approx 1.5$. 
Lastly, we repeated the simulations for a longer loop length, considering the case of a null-point reconnection site, with a total length of $l = 42,000$~km, as discussed in Section~\ref{sec:geom}. In principle, in a longer loop with an asymmetric magnetic field, one would expect a larger time delay of the HXR from the footpoints, due to the extra time that a fraction of the electrons would take to travel from the stronger mirror point to the weaker-field one and precipitating into the dense chromosphere. We find that the time delays of the HXR from the two footpoints is only slightly larger than in the case of the shorter loop: in the order of 0.15~s and 0.22~s for the beam and isotropic pitch-angle distributions, respectively. The other characteristics of this simulation are practically identical Figures~\ref{fig:fokker1} and \ref{fig:fokker2}.

It appears impossible to generate the observed IR time delay and HXR brightness asymmetry self-consistently by using travel-time, trapping or collisional effects. 
However, the foregoing assumes that the IR emission mechanism itself does not involve a delay. Unlike HXR bremsstrahlung, which is prompt, the IR continuum can only appear after sufficient ionization of the chromosphere occurs. The IR flare emission comes from a relatively dense, and normally cool layer \citep{HeinzelAvrett:2012,kaspa2009CEAB...33..309K,simoes2017A&A...605A.125S}, so there are both ionization and recombination timescales implicit in the process. In particular, the increased ionization responsible for the mid-IR emission excesses accumulate, producing a lag relative to the time of actual energy input. This is examined in the next section.

\section{Radiation-hydrodynamic simulations of IR generation}\label{sec:radyn}

A time lag could also be due to the IR formation mechanism, which we investigate with radiation-hydrodynamic simulations. At IR wavelengths, the volume emissivity (photons/cm$^3$/s) is due to the free-free process and depends on the electron and ion number densities, and the electron temperature, i.e.
$\varepsilon_\nu \propto n_e n_i T^{-1/2}$ \citep{OhkiHudson:1975,HeinzelAvrett:2012,simoes2017A&A...605A.125S}.
So different atmospheric conditions in the two footpoints, due to excitation by the different electron beams at the two footpoints, could lead to delays in the growth and decay of free-free emissivity as the ionization level evolves differently. 

The most abundant source of electron-ion pairs is hydrogen ionization, from collisions among the particles in the thermal plasma, from direct non-thermal collisions of the electron beam, or from photoionization. This cannot be calculated straightforwardly since the ionization and recombination rates depend on the atmospheric ionization fraction, temperature and density structure, so that the emissivity will evolve non-linearly as the chromosphere changes due to heating. In the electron-beam heating model, where the beam particles have a range of energies, the fastest electrons will penetrate deep into the chromosphere and mainly result in hydrogen ionization and little heating. The slowest electrons will lose their energy in the upper chromosphere, where the ionization can persist longer. In between these heights, a temperature structure will develop that depends on the beam parameters, the dynamical evolution of the medium, and the radiative transfer. 

To accommodate this complexity, we use radiative hydrodynamic (RHD) simulations that can efficiently model the response of the chromosphere to heating, including evolution in density, temperature and ionisation. 
%

\ifpaper

The radiation hydrodynamics code \textsc{radyn} is a well established tool for investigating chromospheric dynamics \citep{CarlssonStein:1995,1997ApJ...481..500C}, adapted to simulate the chromospheric response to flare energy deposition by an electron beam, including soft X-ray, extreme-UV (EUV) and UV radiation back-warming and photoionisation \citep{AbbettHawley:1999,2005ApJ...630..573A,2015ApJ...809..104A}. \textsc{radyn} solves the plane-parallel, coupled, non-linear equations of hydrodynamics, radiation transfer, charge conservation and atomic level populations on a 1D grid that extends from the sub-photosphere to the corona, representing one side of a symmetric flux tube. The formation and radiative transfer of several spectral lines and continua important for chromospheric energy balance are computed by solving the full non-LTE problem. The ionisation and excitation of the atomic species are computed by considering thermal collisional and radiative rates (including coronal back-warming). For H and He, the effect of direct ionisation by the non-thermal electron beam is also considered, as they are comparable to the thermal rates \citep{1983ApJ...272..739R}{, where we use the treatment of \cite{1993A&A...274..917F} for non-thermal ionisation and excitation of the ground state of hydrogen, and the rates from \cite{1985A&AS...60..425A} for the non-thermal ionisation of He~\textsc{i} and He~\textsc{ii}}.

In a typical flare experiment, an equilibrium pre-flare atmosphere
is heated by an electron beam defined as a power-law with a low energy cutoff $E_c$ in keV, spectral index $\delta$ and the beam intensity $F$ in erg~s$^{-1}$ cm$^{-2}$. The beam transport is treated by self-consistently solving the Fokker-Planck equation in the evolving atmosphere, including the terms for advection, collisional energy losses, collisional pitch-angle scattering, and return currents \citep{2015ApJ...809..104A}. The beam parameters include spectral and pitch-angle distributions as well as the time profile of particle injection.  
We have carried out \textsc{radyn} modeling using as input the electron beam parameters inferred from RHESSI imaging spectroscopy for the northern and southern footpoint sources separately. An estimate of the total energy flux (erg s$^{-1}$ cm$^{-2}$) comes from scaling the total power $P$ by using the normalisation of the photon spectrum of each source, $P_N = K_N/(K_N+K_S)=0.55/(0.55+0.35) \times P \approx 0.6P$ for the northern source, and thus $P_S=K_S/(K_N+K_S)=0.35/(0.55+0.35)\times P \approx 0.4P$ for the southern source. We measured the northern and southern source areas at 30\% and 50\% of the 8.2~$\mu$m image maximum at the time of the peak, obtaining $1.2 \times 10^{17} < A_N < 3.3\times 10^{17}$~cm$^2$, and $5.3 \times 10^{16} < A_S < 1.1 \times 10^{17}$~cm$^2$. Setting $F=P/A$, we have $0.5 \times 10^{10}< F_N < 1.3 \times 10^{10}$~erg~s$^{-1}$ cm$^{-2}$, and $0.9 \times 10^{10}< F_S < 1.9 \times 10^{10}$~erg~s$^{-1}$~cm$^{-2}$ for the northern and southern sources, respectively. These values can still be considered an upper limit for the source areas, as they remain largely unresolved, even by the highest spatial resolution currently available at these or other wavelengths. 

We designed numerical experiments to assess whether the time required for ionizing the hydrogen can cause a delay in the IR emission. For this, after an initial pre-flare period of 1 second, a constant energy flux was injected for 2 seconds, which was then turned off, letting the atmosphere relax for an additional 7 seconds. The output time step of the simulations was set to 0.01 seconds.

For each time, we have the necessary properties of the 1D atmosphere to calculate the IR emission (temperature, electron density, density of the neutral H up to 5 energy levels, and proton density). 
  
From the simulated atmospheric profiles we calculated the electron free-free and H$^-$ free-free opacity coefficients, and radiative transfer at 8.2~$\mu$m, obtaining the time evolution of the brightness temperature $T_B$ \citep{HeinzelAvrett:2012,simoes2017A&A...605A.125S}, shown in Figure \ref{fig:radyn}. As the energy is deposited in the chromosphere, the neutral hydrogen atoms ionise and provide free electrons that increase the optical depth at IR wavelengths. We verified that although He {\sc II} can provide to up 10\% of the free electron content, this occurs higher up in the chromosphere (where the density is lower), and therefore the He contribution of free electrons to IR opacity enhancement is marginal. The rate of increase of the IR brightness temperature varies with the level of energy input, but the formation of the peak in the IR time series depends only on the presence of an energy source: as soon as the energy input is turned off, the hydrogen quickly recombines, depleting the number of free electrons in the chromosphere and reducing the IR emission \citep{simoes2017A&A...605A.125S}. As such, in all the simulated cases, the calculated IR emission has a peak at $t=3$~s, see Figure~\ref{fig:radyn}. We calculated the time delay between the simulated cases, following the same approach described in Section~\ref{sec:IRdata}, and found that it is always smaller than 0.1~s, as shown in Figure~\ref{fig:radyn}c and Figure~\ref{fig:radyn}d. 
\begin{figure*}
\centering
  \includegraphics[width=\textwidth]{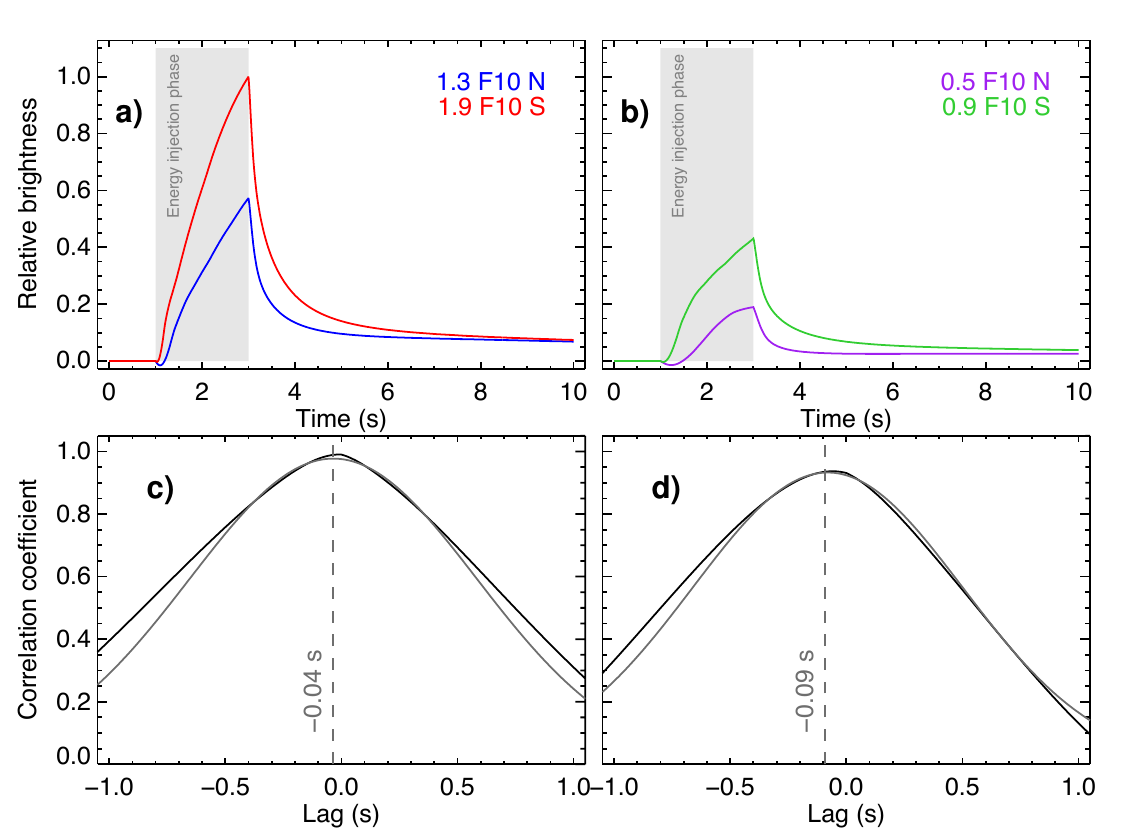}
        \caption{{Results of the RHD simulations} (a) Relative brightness temperature evolution at 8.2~$~\mu$m calculated from the our \textsc{radyn} flaring atmosphere calculations for the upper limit of energy flux injection. (b) Same as panel (a), but for the lower limit of energy input. (c) Correlation coefficient as function of the lag for the simulated footpoint pairs in panel (a) and (d) similar as panel (c), but for the footpoint pairs in panel (b), both showing a lag smaller than 0.1~s.
        \label{fig:radyn}
        }
\end{figure*} 
\fi

In summary, in all simulated cases, as the electron energy flux increases, the resulting IR emission rises monotonically with the hydrogen ionization, providing free electrons which enhance the IR free-free opacity. When the energy input is turned off, the free electrons quickly recombine into neutral hydrogen, reducing the IR emission. 
Cross-correlating the modeled light curves from the two footpoints, we find that there indeed can be a measurable time scale for emission growth, of order 0.05--0.1~s, but this cannot explain the observed lag of 0.75~s, as the timescales of hydrogen ionisation and recombination are too short.

\section{Conclusions}
\label{sec:conclusion}
Our analysis of SOL2014-09-24 has shown the existence of a relative time lag of 0.75\;s in the mid-IR emission from the flare's two footpoint regions. This lag corresponds to a difference in the distances travelled to each footpoint by 30\,keV electrons of $\Delta l = 75,000$\,km apparently longer than can readily be explained by time differences between the arrival of electrons free-streaming to the two chromospheric footpoints. This conclusion includes the possible consequences of the spine-fan configuration revealed by our analysis, and considers
two possible scenarios for forming the IR footpoints, of reconnection at a QSL interior to the fan field (which is favoured by previous authors to explain strong HXR/UV footpoints) and reconnection via the null.

The newer and more precise timing of infrared results {thus} tend to conflict with the earlier observational work \citep{1994PhDT.......335S,1995ApJ...447..923A}, but have smaller observational uncertainties. 
Sophisticated modeling of the delaying effects on the IR emission of coronal magnetic trapping and evolution of chromospheric ionization are also unable to reproduce the observed time delays. Recent observations of double-footpoint structure at 10~$\mu$m (30~THz) in the C-class flare SOL2019-05-20 \citep{lopez2022A&A...657A..51L} also indicate that the double-footpoint signature may result from conductive transport, as well as from non-thermal particle beams.

Our work makes an observational contribution to an ongoing debate on the validity of the long-standing flare electron-beam transport model, by presenting an observation in which the 0.75~s delay between correlated IR signatures at conjugate chromospheric footpoints in a flare {may be} several times longer than can be explained if the energy were transported by fast electron beams streaming through a coronal loop. Supporting numerical simulations eliminate other possible sources of delay that could occur in this model, with electron beam transport simulations predicting a time delay with opposite sign from that observed, and the rapid ionization response of the chromosphere permitting delays of no more than 0.1~s. 
These IR observations are {in tension with} the electron beam model, and what was understood from previous HXR views of impulsive-phase timing. 
We therefore must seek mechanism in addition to electron beams for energy transport from the corona, for example magnetosonic waves \citep{2008ApJ...675.1645F,2012ApJ...756..192L,russel2013ApJ...765...81R,kerr2016ApJ...827..101K,reep2016ApJ...818L..20R,reep2018ApJ...853..101R} {or simple conductive transport from a thermally relaxed coronal plasma.} 
The HXR emission nonetheless requires that significant amounts of flare energy be converted into the kinetic energy of non-thermal electrons, and one implication of our analysis is that -- if they are not significant for energy transport from the corona -- they must be accelerated in the much denser chromosphere, stimulating a new set of theoretical questions.

\section*{Acknowledgements}

We would like to thank the anonymous reviewer for their comments and suggestions, which led us in particular to think seriously about the flare topology. The research leading to these results has received funding from the European Community's Seventh Framework Programme (FP7/2007-2013) under grant agreement no. 606862 (F-CHROMA). PJAS acknowledges support from Conselho Nacional de Desenvolvimento Científico e Tecnológico (CNPq) (contracts 307612/2019-8 and 305808/2022-2), Fundo Mackenzie de Pesquisa e Inovação (MackPesquisa) project 231017 and Fundação de Amparo à Pesquisa do Estado de São Paulo (FAPESP) contracts 2013/24155-3 and 2022/15700-7. LF acknowledges support from grants ST/L000741/1 and ST/X000990/1 made by the UK's Science and Technology Facilities Council. HSH acknowledges the University of Glasgow for support. GSK acknowledges funding via a PhD scholarship from the University of Glasgow's College of Science and Engineering, and from a NASA Early Career Investigator Program award (Grant\# 80NSSC21K0460). The National Solar Observatory is operated by the Association of Universities for Research in Astronomy, Inc. (AURA) under cooperative agreement with the NSF. KFL acknowledges CAPES PROSUC (modality I - 2023) integral financial support.

\section*{Data Availability}

Fokker-Planck code available at \url{https://github.com/pjasimoes/fokker-planck}.
RHESSI data available via the Virtual Solar Observatory (VSO) or Solar Data Analysis Center (SDAC). IR data and simulation results available upon request. \textsc{radyn} code output is available upon reasonable request to the corresponding author. A version of the code is available from: \url{https://folk.universitetetioslo.no/matsc/radyn/} \citep{2023A&A...673A.150C}.


\bibliographystyle{mnras}
\bibliography{IR} 








\bsp	
\label{lastpage}
\end{document}